\documentclass{Rinton-P10x7}
\renewcommand{\be}{\begin{eqnarray}}
\renewcommand{\ee}{\end{eqnarray}}

\begin{document}
\title{Gravitational collapse of a radiating shell and radiation bursts}
\author{Roberto Casadio}
\address{Dipartimento di Fisica, Universit\`a di Bologna,
and I.N.F.N., Sezione di Bologna\\
via Irnerio 46, I-40126 Bologna\\
Email: casadio@bo.infn.it}
\maketitle
\abstracts{
We study the collapse of a self-gravitating thick shell of bosons
coupled to a scalar radiation field.
Due to the non-adiabaticity of the collapse, the shell (quantum)
internal degrees of freedom absorb energy from the (classical)
gravitational field and are excited.
The excitation energy is then emitted in the form of bursts of
(thermal) radiation and the corresponding backreaction on the
trajectory is estimated.}
%
%
%
%
%
%
It is clear that classical physics is not sufficient for a complete
description of the gravitational collapse:
firstly, the collapse ending into a point-like singularity is forbidden
by the uncertainty principle;
secondly, the backreaction of semiclassical effects (with matter evolving
quantum mechanically on a classical space-time) such as the Hawking
radiation~\cite{hawking} should be properly included~\cite{haji}.
We have investigated the semiclassical limit for various
models~\cite{cv,bo} by employing a Born-Oppenheimer
decomposition of the corresponding minisuperspace wavefunction~\cite{bv}.
In particular~\cite{shell}, we considered the collapse in vacuum
of a ``macroshell'' made of $N$ quantized ``microsells'', each
of which corresponding to a typical hadronic mass $m\sim 1\,$GeV $s$-wave
scalar particle, with the proper mass of the macroshell $M=N\,m$.
Due to the bosonic nature of the microshells, the whole system formed a
``condensate'' bound by their mutual gravitational attraction and
it was then shown that the collapse induced non-adiabatic
transitions from the ground state to higher excited states.
Once enough microshells were excited, they could collectively decay back
to the ground state by creating additional particles,
the whole process resulting in a transformation of gravitational energy
into matter.
However, such a model is unrealistic because the absence
of signals from the shell precludes observations,
hence an effective minisuperspace action for radiating
shells was derived~\cite{vaidya} as a key tool for making observational
predictions~\cite{qg99,next}.
\par
We order the microshells according to their area ($r_1<r_2<\ldots<r_N$)
and assume the thickness of the macroshell $\delta=r_N-r_1$
is small ($\delta\ll r_1$).
The space between two microshells has Schwarzschild geometry with
mass function $M_i$ which vanishes for $r<r_1$ and equals the total ADM
mass $M_N\equiv M_s$ for $r>r_N$.
We also assume the microshells start out at large radii with negligible
velocity, hence $M_{i+1}=M_i+m$, and use the standard junction
equations~\cite{israel}.
It is then convenient to consider each microshell in the mean field
of the others (Hartree approximation~\cite{messiah,shell}) which yields
a classical effective Hamiltonian for each microshell~\cite{next}
$H_m=({1/2})\,m\,\dot{\bar r}^2+V_m$,
where an overdot denotes the derivative with respect to the microshell
proper time $\tau$, $\bar r=r-R$ is the displacement of the microshell
relative to the average radius $R$ and, to leading order in
$|\bar r|/R$ and $m/M$,
\be
V_m\simeq
\left\{\begin{array}{ll}
{G\,M_s\,m\over 2\,R}\,{\bar r\over R}
\,\left(1-{G\,M^2\over 2\,R\,M_s}
\right)
&\ \ \ \ \ \ \ \ \ \
\bar r>+{\delta\over 2}
\\
{G\,M_s\,m\over 2\,R^2}\,
\left({\bar r^2\over \delta}+{\delta\over 4}\right)
-{G^2\,M^2\,m\over 4\,R^3}\,\bar r
& \ \ \ \ \ \ \ \ \ \
|\bar r|\le {\delta\over 2}
\\
{G\,M_s\,m\over 2\,R}\,{\bar r\over R}\,
\left(-1-{G\,M^2\over 2\,R\,M_s}
\right)
&\ \ \ \ \ \ \ \ \ \
\bar r<-{\delta\over 2}\ .
\end{array}
\right.
\label{Vm}
\ee
The potential $V_m$ accounts for tidal effects (backreaction) between the
microshells and, for $R>R_H\equiv 2GM_s$, confines the microshells
around $\bar r=0$ within the thickness~\cite{shell}
\be
\delta\sim\ell_m^{2/3}\,R^{1/3}\,\left({R\over R_H}\right)^{1/3}
\ ,
\label{d}
\ee
where $\ell_m=\hbar/m$ is the Compton wavelength of a microshell.
\par
We can now quantize the microshells and obtain a Schr\"odinger
equation with explicit time-dependence due to $R=R(\tau)$ in the
potential,
\be
i\,\hbar\,\dot\Phi=\left[{\hat\pi_r^2\over 2\,m}+\hat V_m\right]
\,\Phi
\ .
\label{Schrod}
\ee
One can solve this equation by making use of invariant
operators \cite{lewis-ries} and then compute the transition
amplitudes $A_{0\to n}(\tau)$ from the ground state
$\Phi_0(0)$ to a state $\Phi_n(\tau)$ with (higher) energy
$E_{n}=n\,\hbar\,\Omega$, where
$\Omega=(1/R)\,\sqrt{{R_H/2\,\delta}}$.
Let us also note that, since the widths of the lowest states are
of order $\delta$, the bosonic microshells
are essentially superimposed and form a condensate~\cite{next}.
To lowest order in $\dot R$, one finds $A_{0\to 0}\simeq 1$,
$A_{0\to 2\,n+1}=0$ and
\be
A_{0\to 2\,n}(\tau)\simeq
(-i)^n\,{\sqrt{(2\,n)!}\over 3^{n}\,2^{n/2}\,n!}\,
\left({\delta\over R_H}\right)^{n/2}\,\dot R^{n}
\ .
\label{0_2n}
\ee
We remark that the above transition amplitude is a consequence
of both a quantum mechanical non-adiabatic effect
($A_{0\to 2\,n}\propto \dot R^{n}$) and the finite thickness
of the macroshell ($A_{0\to 2\,n}\propto \delta^{n/2}$),
the latter being further related to the quantum mechanical
nature of the model ($\delta\propto \hbar^{2/3}$).
The above expression is, for realistic cases, small, and
the probability for a microshell to get excited twice during
the whole collapse is therefore negligible.
\par
Since the proper mass of each microshell has a non-vanishing
probaility to increase in time, the collapse is not a free fall:
the trajectory of the macroshell slows down~\cite{shell},
and this effect is further sustained by the emission~\cite{vaidya}
which quickly brings the proper mass back to the initial value.
We shall then consider $M$ as effectively constant along the
collapse and compute the net variation of $M_s$ due to the
loss of the (excited state) proper energy in time.
The whole process is a transformation of gravitational
energy into radiation.
Of course, if $\dot R=0$ for $R\sim R_H$, the proper acceleration
of the macroshell would equal the surface gravity $a_H=1/2\,R_H$
of a black hole of mass $M_s$.
In this limiting case one could exploit the analogy between
Rindler coordinates for an accelerated observer in flat
space-time and Schwarzschild coordinates for static observers
in a black hole background \cite{birrell} and find that the
shell emits the excess energy with Hawking temperature
\be
T_H={\hbar\,a_H\over 2\,\pi\,k_B}={\hbar/k_B\over 8\,\pi\,M_s}
\ ,
\label{T_H}
\ee
where $k_B$ is the Boltzmann constant.
We previously proposed an explicit construction which leads to
this result~\cite{cv}, but here we shall determine $R=R(\tau)$
from the equation of motion.
\par
We now consider an isotropic massless scalar field $\varphi$
conformally coupled to gravity~\cite{birrell} and to the microshells
via the radiation coupling constant $e$, and study the emission
of quanta of radiation occurring when the microshells in the state
$\Phi_{2}$ decay back into the ground state $\Phi_0$.
For the cases we consider (see Table~\ref{table1} for an example)
the emissions occur practically in phase, since the ratio between
the thickness of the shell and the typical wavelength of the emitted
quanta is small and the distance covered between each of the numerous
emissions is small, again with respect to the radiation wavelength.
This process is therefore coherent and the transition probability
per unit Schwarzschild time ($dP/dt$) can be easily obtained in
perturbation theory to leading order in $e$~\cite{next}.
\par
It is then straightforward to estimate the rate of proper energy
lost by the macroshell per unit (Schwarzschild) time,
\be
{dE\over dt}=-4\,\pi\,R^2\,
\sum_\omega\,\mu(\omega)\,\Gamma(\omega)\,\hbar\,\omega\,
{dP\over dt}
\ ,
\label{lum}
\ee
where $\mu(\omega)=(1-R_H/R)^{3/2}\,\omega^2$ the phase space measure
for radiation quanta of frequency $\omega$ (measured at the shell
position $r=R$) and $\Gamma\sim 1$ the gray-body factor for zero
angular momentum outgoing scalar waves.
The sum in Eq.~(\ref{lum}) is dominated by the contribution with
$\omega=2\,\Omega$ and the flux becomes
\be
{dE\over dt}&\simeq&
-{16\,\pi^2\,e^2\over 9}\,
{N^2\,\ell_m^{8/3}\over R_{\phantom H}^{10/3}\,R_H^{4/3}}\,
\left(1-{R_H\over R}\right)\,
{\dot R^2\over e^{2\,\Omega\,\sqrt{1-R_H/R}/k_B\,T_H}-1}
\ ,
\label{flux}
\ee
in which there appears a Planckian factor with Tolman shifted
(instantaneous) Hawking temperature.
We note that this feature is not sufficient to relate
our model to the usual Hawking effect, since Eq.~(\ref{flux}) contains a
coupling constant $e$, while the probability of Hawking
emission does not depend on any coupling~\cite{hawking}.
\par
\begin{figure}
\centering
\raisebox{3.3cm}{$R$}\hspace{-0.2cm}
\epsfxsize=2.5in
\epsfbox{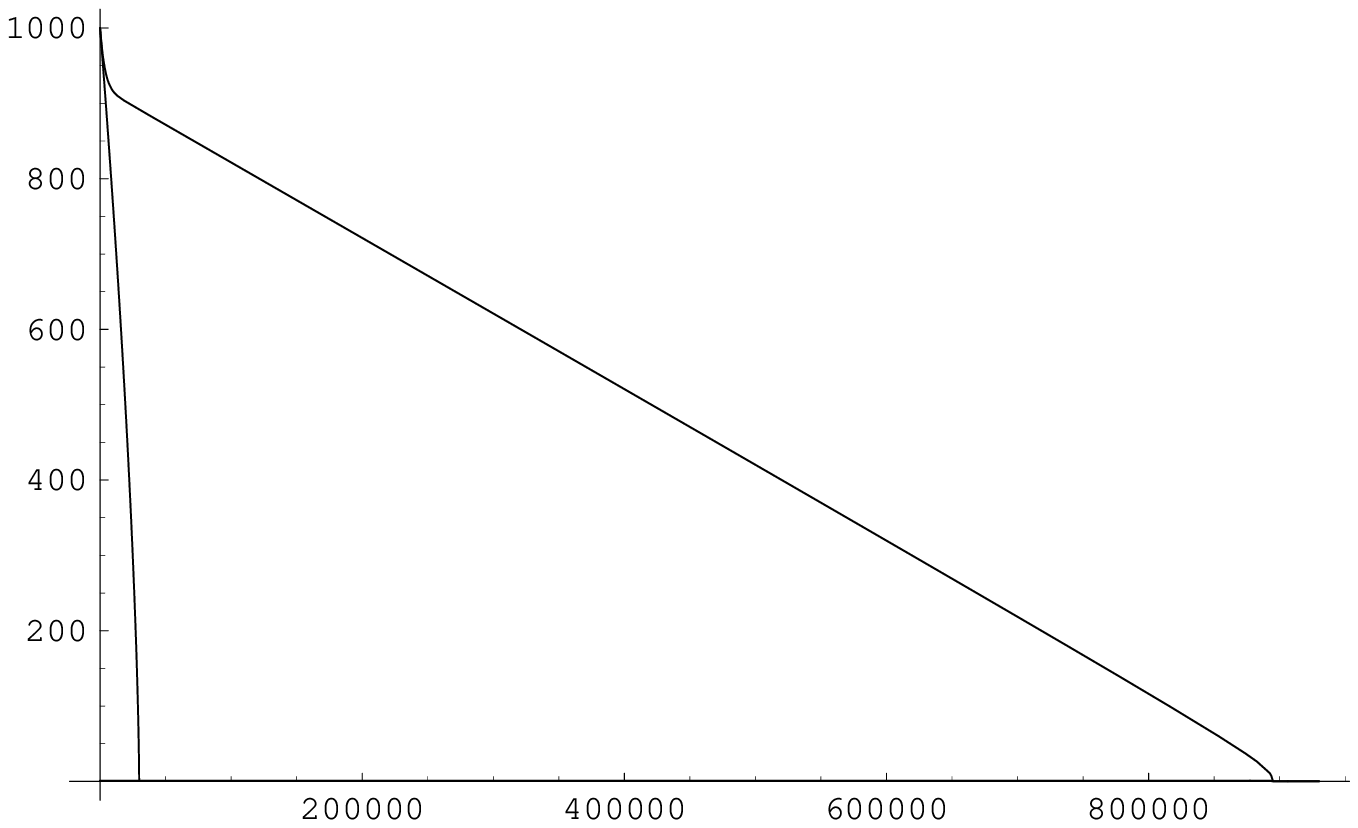}
%
\raisebox{3.3cm}{$\dot R$}\hspace{-0.2cm}
\epsfxsize=2.5in
\epsfbox{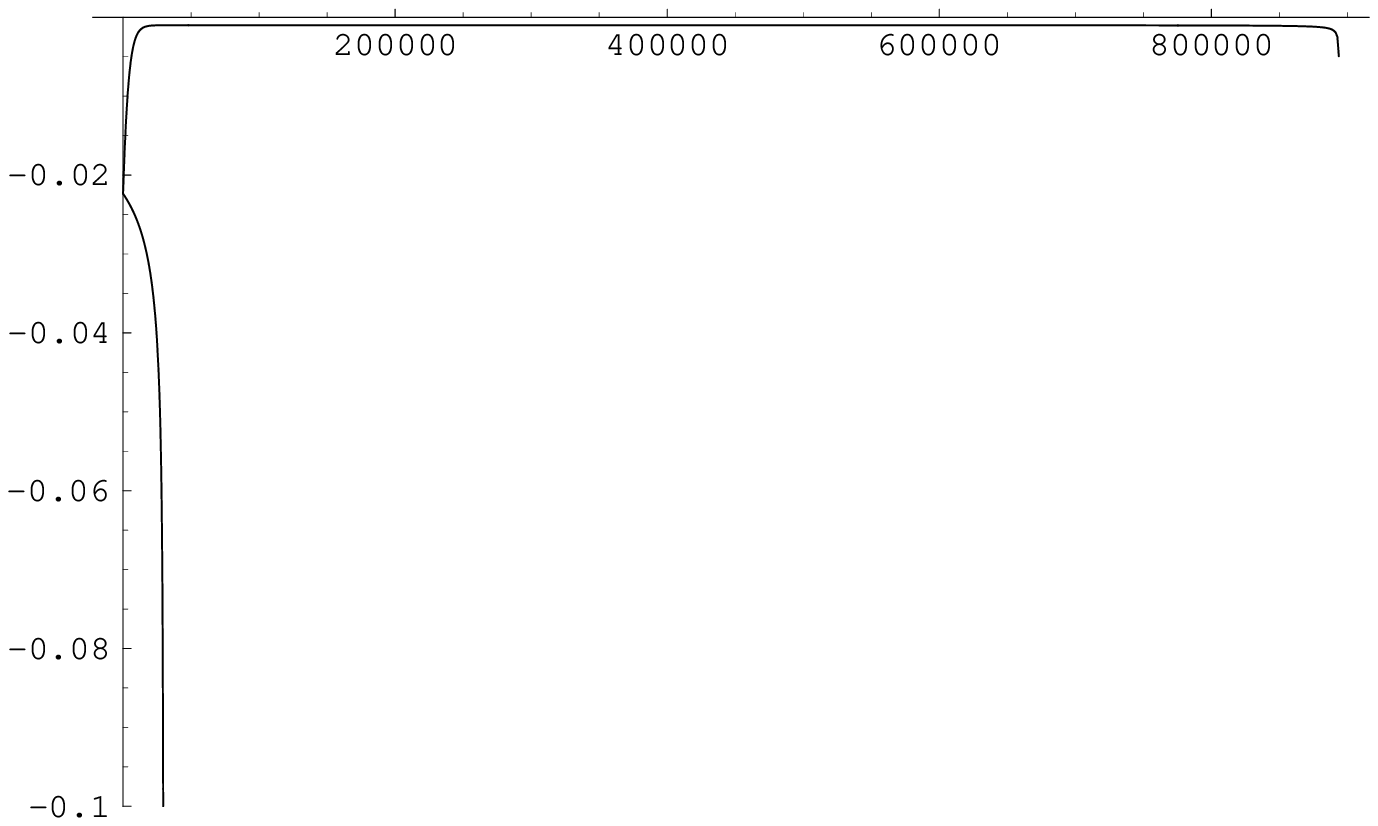}
\caption{Trajectory $R(\tau)$ and velocity $\dot R(\tau)$ of the radiating
shell in units of $R_H(0)=4\cdot 10^{-12}\,$cm for $N=2\cdot 10^{40}$ and
$e=8\cdot 10^{-16}$ (upper curves) compared to the non-radiating collapse
(lower curves).
}
\label{RN2E40}
\end{figure}

One can now integrate numerically the equation of motion for the
(average) radius of the macroshell \cite{israel},
\be
\dot R^2=-1+\left({M_s\over M}\right)^2+{G\,M_s\over R}
+{G^2\,M^2\over 4\,R^2}
\ ,
\label{eqn1}
\ee
together with the equation for $M_s$ \cite{vaidya},
\be
\dot M_s\simeq {dE\over dt}
\ ,
\label{eqn2}
\ee
until $R$ is larger than a few times $R_H$ (since our approximations
break down for $R\sim R_H$~\cite{shell,next}).
In general, one finds that the non-adiabatic excitations giving appreciable
changes in $M_s$ occur relatively close to the horizon and there is no
strong dependence on the initial value of $R$.
However, despite the small change in $M_s$ for $R\gg R_H$, we find
a large backreaction on the trajectory~\cite{next} (see Table~\ref{table1}
and Fig.~\ref{RN2E40} for a typical case).
In particular $\dot R$ remains (negative and) small (mostly
within a fraction of a percent of the speed of light), thus confirming
our approximation scheme in which we just kept the leading order in
$\dot R$.
\par
To conclude we have seen that the states in a collapsing shell of
bosonic matter, initially in free fall, become excited due to
quantum non-adiabatic transitions and gravitational energy is lost
through the emission of scalar radiation.
This effect is intrinsically quantum mechanical, since it is a
consequence of the quantum mechanical--bound state nature of the
macroshell and the coherence of the emitted radiation.
This can cause the shell to lose enough energy so that the
backreaction on the trajectory of the radius is large.
Since the origin of the studied effect is the non-adiabaticity
of the collapse and coherence, one might argue that an analogous
phenomenon can happen for all collapsing matter, including the
accretion disks around black holes.
Our result would thus suggest a new mechanism by which
the accreting matter can emit radiation.
Whether this mechanism can be related to the Hawking effect
is a point which requires further study.
\begin{table}
\centering
\begin{tabular}{|c|c|}
\hline
radiation wavelength $\lambda$ & $10^{-10}\div 10^{-7}\,$cm \\
\hline
$\delta/\lambda$ & $10^{-3}\div 10^{-1}$ \\
\hline
total number of emissions & $10^{39}$ \\
\hline
$N_\delta$ & $10^{35}\div 10^{36}$ \\
\hline
\end{tabular}
\caption{
Typical values of the relevant quantities for the case in
Fig.~\ref{RN2E40}.
$N_\delta$ is the average number of emissions while the shell
moves a space $\delta$.}
\label{table1}
\end{table}
\end{document}